\documentclass[pra,aps,showpacs,twocolumn,floatfix]{revtex4-1}
\usepackage{graphicx}
\usepackage[ansinew]{inputenc}
\usepackage{array}
\usepackage{color}
\usepackage{amsmath}
\usepackage{amsxtra}
\usepackage{amstext}
\usepackage{amssymb}
\usepackage{latexsym}
\usepackage{dsfont}
\begin{document}

\title{Multimode strong-coupling quantum optomechanics}

\author{H.~Seok, L.~F.~Buchmann, E.~M.~Wright, P.~Meystre}
\affiliation{B2 Institute, Department of Physics and College of Optical Sciences \\
The University of Arizona, Tucson, Arizona 85721}

\begin{abstract}
We study theoretically the dynamics of multiple mechanical oscillators coupled to a single cavity field mode via linear or quadratic optomechanical interactions. We focus specifically on the strong coupling regime where the cavity decays much faster than the mechanical modes, and the optomechanical coupling is comparable to or larger than the mechanical frequency, so that both the optical and mechanical systems operate in the deep quantum regime.  Using the examples of one and two mechanical oscillators we show that the system can classically exhibit bistability and bifurcations, and we explore how these manifest themselves in interference, entanglement, and correlation in the quantum theory, while revealing the impact of decoherence of the mechanical system due to cavity fluctuations and coherent driving.   
\end{abstract}

\maketitle

\section{Introduction}

Cavity optomechanics has undergone tremendous growth owing to recent advances of nano- and micro-fabrication, culminating in the cooling of macroscopic mechanical objects to near the ground state~\cite{cooling1, cooling2, cooling3, cooling4}.  Further developments promise a variety of applications, for example  precision measurements of feeble forces, masses and displacements ~\cite{force1, force2, force3}, coherent control of quantum states for quantum information science~\cite{information1, information2, information3} and fundamental tests of quantum mechanics with macroscopic objects~\cite{fundamental1, fundamental2, fundamental3}, to mention just a few. 

The linear optomechanical coupling of a cavity field to the position of a mechanical mode is typically modeled using a Fabry-P{\' e}rot cavity with a moving end mirror, and allows one to elucidate such effects as optical bistability, optical spring effects, and radiation-induced cooling of the mechanical oscillators ~\cite{review0, review1, review2, review3, review4, review5, review6}. Alternative experimental setups such as the membrane-in-the-middle geometry ~\cite{membrane0,membrane1, membrane2},  cold atoms in a cavity ~\cite{atom1, atom2, atom3, atom4},  and photonic crystal structures ~\cite{cooling3, cooling4}, facilitate the investigation of quadratic or higher-order optomechanical interactions~\cite{quadratic1, quadratic2}, and opens up the exploration of multimode optomechanics~\cite{multimode0, multimode1, multimode2, multimode3, multimode4, multimode5, multimode6, multimode7}.

Most experiments to date are carried out in a regime where the optomechanical coupling is weak compared to the mechanical frequency and the cavity linewidth, and can be treated as a perturbation. When the cavity is driven by a classical field, the optomechanical coupling in this regime can be linearized, resulting in effects such as beam splitter swapping and parametric amplification. This linearized optomechanical interaction also accounts e.g. for radiation-induced cooling or amplification~\cite{review1, review2, review3, review4, review5, review6}, normal-mode splitting~\cite{mode_splitting1}, and optomechanically induced transparency~\cite{OMIT1, OMIT2, OMIT3}.

However, the linearized treatment needs to be revisited if the optomechanical interaction frequency becomes comparable to the frequency of harmonic trapping potential of the mechanics, in which case its intrinsic nonlinearity becomes of crucial importance even for cavity fields that only contain very few photons~\cite{blockade}. Here the role of quantum fluctuations can become of central importance. The so realized nonlinear optomechanics promises the generation of non-Gaussian states~\cite{non-Gaussian},  nonclassical mechanical steady-state~\cite{nonclassical_mechanics}, and quantum state control and optical switching at a single photon level~\cite{nonclassical}. Such a strong-coupling regime has not been realized in nano-fabricated optomechanical systems, but optomechanical systems involving ultracold atomic clouds~\cite{atom1, atom2, atom3, atom4} can currently operate in this regime. 

In this paper, we study theoretically optomechanical interactions in the single-photon strong-coupling regime in which the optomechanical coupling is comparable to or stronger than the harmonic trapping potential of the mechanics, and cavity dissipation is the dominant source of damping. We first investigate the classical effective potential of the mechanical oscillators coupled to a single mode cavity field via a strong linear or quadratic coupling. The classical theory is used to elucidate selected examples that are then explored in the quantum theory.  A quantum master equation is solved numerically to explore the dynamics of the mechanical oscillators and the effects of cavity fluctuations on the mechanics for the cases of one or two mechanical oscillators. We demonstrate that for linear coupling a single mechanical oscillator in the deep quantum regime does not exhibit the bistability that exists in the classical regime, a result that parallels the situation familiar from cavity QED~\cite{QED_bistability1, QED_bistability2}. In the quadratic coupling case the cavity fields leads to splitting and recombination dynamics of the spatial wave-functions of the oscillators, realizing a quantum interferometer. We also demonstrate the contribution of cavity fluctuations on the coherence of the mechanical system and propose a way to increase the lifetime of that coherence. 

The remainder of this paper is organized as follows. Section II presents our basic model, introducing both linear and quadratic coupling of multimode mechanical systems to a single-mode optical field.  Section III reviews how the adiabatic elimination of the optical field results in effective nonlinear dynamics for the mechanics and discusses its most important properties in the classical limit, concentrating of the form of the effective potential governing this dynamics for both linear and quadratic optomechanical couplings, and in the single-mode and two-mode cases. Section IV addresses the extension of the treatment to the quantum regime and discusses the additional multiplicative quantum noise resulting from the elimination of the optical field.  Section V then presents a number of examples for one and two mechanical oscillators and explores how quantum fluctuations manifest themselves in interference, entanglement, and correlations in the quantum theory, while revealing the impact of decoherence of the mechanical system due to cavity fluctuations.  Finally Section VI is a summary and outlook.

\section{Basic Model}\label{Basic}
In this section we introduce the quantized Hamiltonian for our system and evaluate the corresponding Heisenberg-Langevin equations of motion.  We consider an optomechanical system composed of $\cal N$ identical mechanical oscillators, with effective mass $m$ and frequency $\omega_m$, that are coupled via optomechanical interactions to a single cavity mode.  The Hamiltonian describing the system is
\begin{equation}
H = H_{\rm opt} + H_{\rm m} + H_{\rm om}+ H_{\rm loss}  ,
\end{equation} 
where the cavity field Hamiltonian 
\begin{equation}
H_{\rm opt} = \hbar\omega_c\hat{a}^{\dag}\hat{a}+ i\hbar(\eta e^{-i\omega_L t}\hat{a}^{\dag}-\eta^{*} e^{i\omega_L t}\hat{a}) ,
\end{equation}
describes the cavity mode of frequency $\omega_c$ driven by an external field of frequency $\omega_L$ with pumping parameter $\eta$, and
\begin{equation}
H_{\rm m} = \frac{\hbar\omega_m}{2}\sum_{k=1}^{\cal N}(\hat{p}_k^2+\hat{x}_k^2) 
\end{equation}
is the mechanical Hamiltonian for the $\mathcal{N}$ identical modes, $\hat{x}_k$ and  $\hat{p}_k$ being the dimensionless position and momentum operators for the $k$-th mirror, respectively. Here the dimensionless position and momentum operators can be obtained as their dimensional counterparts in units of 
\begin{equation}
x_{0} = \sqrt{\hbar/(m\omega_m)}, \quad p_0 = \sqrt{m\hbar\omega_m} ,
\end{equation}
 so that 
 \begin{equation}
 [\hat{x}_j, \hat{p}_k] = i\delta_{jk}.
 \end{equation}
Linear optomechanical interactions are described by the Hamiltonian
\begin{equation}
H_{\rm om} = -\hbar\hat{a}^{\dag}\hat{a}\sum_{k}^{\cal N} g_{0, k}\hat{x}_k \label{linear},
\end{equation}
whereas quadratic optomechanical interactions are accounted for using
\begin{equation}
H_{\rm om} = \hbar \hat{a}^{\dag}\hat{a}\sum_{k}^{\cal N}g_{0, k}^{(2)}\hat{x}_k^2 \label{quadratic}, 
\end{equation}
where $g_{0, k}$ and $g_{0, k}^{(2)}$ are the linear and quadratic single-photon coupling coefficients, respectively.   Finally, $H_{\rm loss}$ describes the interaction of the cavity field and the mechanical modes with their respective reservoirs and accounts for dissipation.

For both linear and quadratic optomechanical interactions the Heisenberg-Langevin equations of motion for the cavity and mechanical modes may be evaluated using the standard input-output formalism~\cite{inputoutput}, and we adopt a frame rotating with the laser frequency $\omega_L$.  For the case of linear interactions described by the Hamiltonian (\ref{linear}) the operator Heisenberg-Langevin equations are
\begin{eqnarray}
\dot{\hat{x}}_j &=& \omega_m\hat{p}_j, \\
\dot{\hat{p}}_j &=& -\omega_m\hat{x}_j+g_{0, j}\hat{a}^{\dag}\hat{a}
-\frac{\gamma}{2}\hat{p}_j+\hat{\xi}, \label{linear_p} \\
\dot{\hat{a}} &=& i\left[\Delta_c +\sum_{k}^{\cal N}g_{0, k}\hat{x}_k\right]\hat{a}-\frac{\kappa}{2}\hat{a}+\eta +\sqrt{\kappa}\hat{a}_{\rm in} ,
\label{linear_q_cavity}
\end{eqnarray}
where $\Delta_c = \omega_L-\omega_c$ is the detuning between the pump and cavity frequencies, and $\kappa(\gamma)$ and $\hat{a}_{\rm in}(\hat{\xi})$ are the cavity (mechanics) decay rate and corresponding noise operator.   In a similar manner for the case of quadratic interactions described by the Hamiltonian (\ref{quadratic}) the Heisenberg-Langevin equations are given by
\begin{eqnarray}
\dot{\hat{x}}_j &=& \omega_m\hat{p}_j, \\
\dot{\hat{p}}_j &=& -(\omega_m+2g_{0, j}^{(2)}\hat{a}^{\dag}\hat{a})\hat{x}_j-\frac{\gamma}{2}\hat{p}_j+\hat{\xi}, \label{p} \\
\dot{\hat{a}} &=& i\left[\Delta_c -\sum_{k}^{\cal N}g_{0, k}^{(2)}\hat{x}_k^2\right]\hat{a}-\frac{\kappa}{2}\hat{a}+\eta +\sqrt{\kappa}\hat{a}_{\rm in} \label{q_cavity}.
\end{eqnarray}
This completes the description of our basic model. 

\section{Classical theory}\label{classical}

We first outline aspects of the classical theory that will be useful to frame the results of the quantum theory discussed in the next sections.  We specifically consider the regime in which the cavity decay rate is much larger than all other system rates, including the mechanical frequency and decay rate, and the single-photon optomechanical coupling coefficients.   In this regime we derive effective potentials for the mechanics for the cases of both linear and quadratic interactions. These allow us to identify interesting operating conditions for each case.  

\subsection{Linear Interactions}
The classical theory applies when the cavity field and mechanical modes are sufficiently highly excited that fluctuations around their mean-field values may be neglected.  In this limit the quantum operators may be replaced by their $c$-number expectation values, $\hat{a} \rightarrow \alpha,~\hat{x}_j \rightarrow x_j,~\hat{p}_j \rightarrow p_j$. For the case of linear interactions this leads to the mean-field equations
\begin{eqnarray}
\dot{x}_j &=& \omega_m p_j, \label{linear_cx} \\
\dot{p}_j &=& -\omega_m x_j+g_{0, j}|\alpha|^2 -\frac{\gamma}{2} p_j,  \label{linear_cp}\\
\dot{\alpha} &=& i\left[\Delta_c +\sum_{k}^{\cal N}g_{0, k} x_k\right]\alpha-\frac{\kappa}{2}\alpha+\eta.
\end{eqnarray}
If the cavity decay rate $\kappa$ is much larger than the decay rate of the mechanical modes and the coupling strengths, $\kappa \gg \{\gamma,~g_{0, j}\}$, the cavity field may be adiabatically eliminated on time scales greater than $1/\kappa$ to yield
\begin{equation}
\alpha (t) \approx \frac{\eta}{-i[\Delta_c+\sum_{k}^{\cal N}g_{0, k} x_k(t)]+\kappa/2} \label{linear_cavity}. 
\end{equation}
Substituting Eq.~(\ref{linear_cavity}) into Eq.~(\ref{linear_cp}) then gives 
\begin{equation}
\dot{p}_j = -\omega_m x_j +\frac{g_{0, j}|\eta|^2}{[\Delta_c+\sum_{k}^{\cal N}g_{0, k} x_k]^2+{\kappa^2\over 4}} -\frac{\gamma}{2}p_j, \label{linear_cforce}
\end{equation}
where the first term of the right-hand-side is the harmonic restoring force for the mechanics, the second term describes the radiation pressure force due to the linear optomechanical coupling, and the last term is the mechanical damping force. Note that the mechanical modes are coupled through the radiation pressure force from the shared optical field. 

Solving Eq.~(\ref{linear_cforce}) with $\dot{p}_j=0$ yields a set of ${\cal N}$ coupled equations for the mechanical mode positions $x_{j, {\rm s}}$ in steady-state
\begin{equation}
\left[\left(\Delta_c+\sum_{k}^{\cal N}g_{0, k} x_{k, {\rm s}}\right)^2+\frac{\kappa^2}{4}\right]x_{j, {\rm s}} = \frac{g_{0, j}|\eta|^2}{\omega_m}  \label{linear_x_s}.
\end{equation}
For the case that the coupling coefficients are identical, $g_{0, k}\equiv g_0$,  the steady-state positions of the mechanical modes must also be identical $x_{j, {\rm s}}=x_s$, since the quantity in the bracket in Eq.~(\ref{linear_x_s}) has the same value for all mechanical modes, as does the right hand side of the equation.  Here $x_s$ satisfies the cubic polynomial equation
\begin{equation}
 g_0^2 {\cal N}^2 x_s^3 + 2 \Delta_c  g_0 {\cal N} x_s^2 + \left[\Delta_c^2+\frac{\kappa^2}{4}\right] x_s - \frac{g_0 |\eta|^2}{\omega_m} = 0 \label{cubic}.
\end{equation}
The number of physical solutions for the steady-state position $x_s$ depends on the discriminant of Eq.~(\ref{cubic}).   For a positive discriminant there can be three solutions, with possible multistability, whereas a single solution results for a negative discriminant.  Since the formula for the discriminant is rather long and complicated we make use of an alternative procedure to assess the number of the physical solutions: A necessary condition for the discriminant in Eq.~(\ref{cubic}) to be positive is that the first derivative of the equation with respect to $x_s$ should have two distinct roots, yielding the condition
\begin{equation}
{|\Delta_c|\over\kappa} > \frac{\sqrt{3}}{2}. 
\label{multiple}
\end{equation}
When this condition is satisfied the mechanics can exhibit three steady-state solutions for appropriate values of the cavity pumping rate $|\eta|$. 

The stability of the steady-state position of the mechanics can be investigated by linearizing the equations of motion ~(\ref{linear_cx}) and  ~(\ref{linear_cforce}) for small mechanical fluctuations. Here we employ the alternative approach of examining the effective potential $U_{\rm eff}$ for the mechanics in terms of which Eq. (\ref{linear_cforce}) can be written as $\dot{p}_j=-\frac{1}{\hbar}{\partial U_{\rm eff}\over \partial x_j}$ in the absence of dissipation, where
 \begin{eqnarray}
U_{\rm eff} &=& \frac{\hbar\omega_m}{2}\sum_{k}^{\cal N}x_{k}^2 \nonumber \\
&-&\frac{2\hbar|\eta|^2}{\kappa}\arctan\left[\frac{\Delta_c+\sum_{k}^{\cal N} g_{0_k} x_{k}}{\kappa/2}\right] .
\label{linear_cpotential}
\end{eqnarray}
We next present two examples of the effective potential to highlight interesting features of the classical model with linear interactions, with a view to exploring these further in the quantum theory.
 
\begin{figure}[]
\includegraphics[width=0.48 \textwidth]{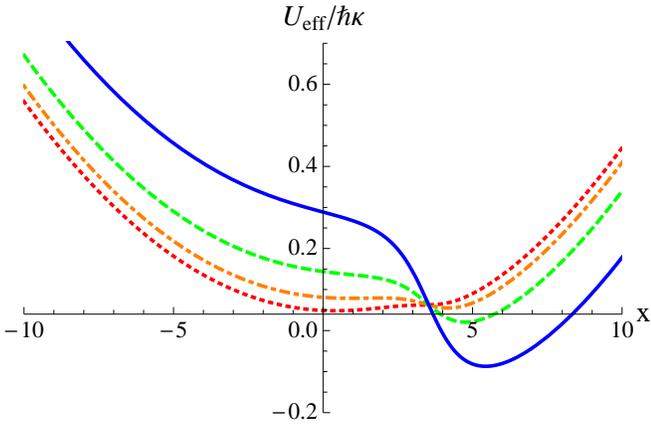}
\caption{\label{fig:linear_single_potential} (Color online) Effective potential $U_{\rm eff}(x)$, in units of $\hbar\kappa$, versus the dimensional position $x$ for a single mechanical oscillator linearly coupled to a cavity mode, and for the several values of the normalized cavity pumping rate: $|\eta|/\kappa = 0.14$ (red dotted line), $|\eta|/\kappa = 0.18$ (orange dot-dash line), $|\eta|/\kappa = 0.24$ (green dashed line), $|\eta|/\kappa = 0.34$ (blue solid line). The potential with $|\eta|/\kappa= 0.18$ exhibits two local minima corresponding to stable solutions and one local maximum corresponding to an unstable solution. Here $~\omega_m/\kappa = 0.01, ~g_{0, 1}/\kappa =0.3,~\Delta_c/\kappa = -1.5$.  }
\end{figure}

As a first example Fig.~\ref{fig:linear_single_potential} shows the effective potential $U_{\rm eff}(x)$ as a function of the dimensionless mechanical position $x$ for a single mode, ${\cal N} =1$, and for a variety of values of the normalized cavity pumping rate $|\eta|/\kappa$, with $\kappa$ the field decay rate.  The ratio of the cavity detuning to the decay rate is chosen as  $\Delta_c/\kappa = -1.5$, meaning that the condition in Eq. (\ref{multiple}) is satisfied and multiple solutions are possible, other fixed parameters being given in the figure caption. Here and in all the following figures, we use dimensionless positions.  

What Fig.~\ref{fig:linear_single_potential} illustrates is that for the lower values of the cavity pumping rate the effective potential has a single stable minimum, the dimensionless position $x$ of the minimum increasing monotonically from zero with increasing pumping rate.  For large enough pumping rates, however, a double-well effective potential arises with two stable minima and one unstable maximum (see the orange dot-dash curve for $|\eta|/\kappa = 0.18$), indicating bistability in the mechanical response.  For still larger pump rates the effective potential again exhibits a single stable minimum position displaced from the origin.  Thus, as is well known, even for a single mechanical mode and linear interactions a bistable mechanical response can arise~\cite{classical_bistability1, membrane0}. We return to this example in Sec.~\ref{ResultsLin} to discuss the impact of quantum fluctuations on that behavior. 

\begin{figure}[]
\includegraphics[width=0.48 \textwidth]{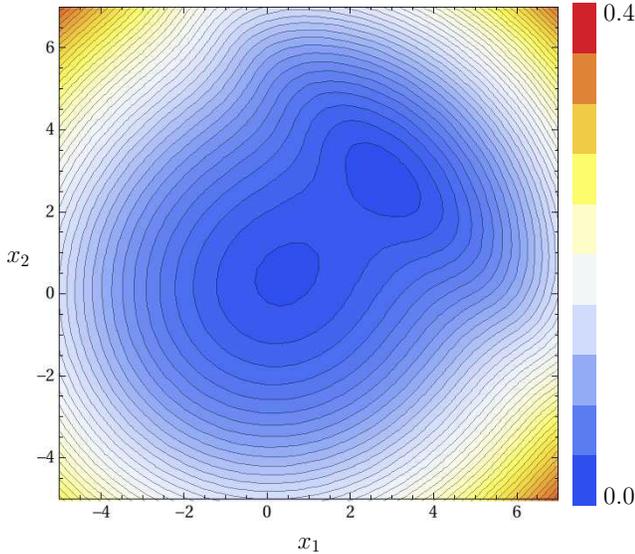}
\caption{\label{fig:lin_two_pot}   (Color online) Effective potential $U_{\rm eff}(x_1,x_2)$, in units of $\hbar\kappa$, as a function of the dimensionless positions $x_{1}$ and $x_2$ for two mechanical oscillators linearly coupled to the cavity mode. Here $~\omega_m/\kappa = 0.01, ~g_{0, 1}/\kappa =g_{0, 2}/\kappa =0.3,~\Delta_c/\kappa = -1.5,~|\eta|/\kappa = 0.16$.}
\end{figure}

As a second example we consider the case of two mechanical modes with linear interactions and with equal values of the coupling strengths, $g_{0, 1}=g_{0, 2}$.  Figure \ref{fig:lin_two_pot} shows the effective potential $U_{\rm eff}(x_1,x_2)$ versus the dimensionless positions $x_{1}$ and $x_2$ of the two modes for the parameters given in the figure caption.  What is interesting about this case is that there are two stable minima of equal depth, situated on the line $x_1=x_2$ by virtue of the equal coupling constants.  We shall explore the possibility that in the quantum regime this effective potential can lead to entangled quantum states in Sec.~\ref{ResultsLin}.

\subsection{Quadratic Interactions}

Following the same procedure of replacing the quantum operators with their $c$-number expectation values to realize the classical theory, we find the following mean-field equations of motion for the case of quadratic interactions
\begin{eqnarray}
\dot{{x}}_j &=& \omega_m {p}_j, \label{xclassical} \\
\dot{{p}}_j &=& -(\omega_m+2g_{0, j}^{(2)}|\alpha|^2){x}_j- \frac{\gamma}{2} {p}_j, \label{pclassical}\\
\dot{\alpha} &=& i\left[\Delta_c-\sum_{k}^{\cal N} g_{0, k}^{(2)} x_k^2\right]\alpha-\frac{\kappa}{2}\alpha+\eta.
\end{eqnarray} 
Then upon adiabatically eliminating the cavity mode field as before we obtain the equations of motion for the mechanical modes 
\begin{equation}
\dot{p}_j = -\omega_m x_j-\frac{2g_{0, j}^{(2)}|\eta|^2}{[\Delta_c- \sum_{k=1}^{\cal N} g_{0, k}^{(2)} x_k^2]^2+{\kappa^2\over 4}}{x}_j-\frac{\gamma}{2}p_j \label{cforce}.
\end{equation}
In contrast to the case of linear interactions the radiation pressure force due to the quadratic interactions does not displace the mechanical oscillators, but rather shifts their mechanical frequencies, as is well known.  

Solving Eq. (\ref{cforce}) with $\dot{p}_j =0$ yields for the steady-state displacements $x_{j, {\rm s}}$ of the mechanical modes 
\begin{eqnarray}
\left[\omega_m+\frac{2g_{0, j}^{(2)}|\eta|^2}{\left(\Delta_c-\sum_{k}^{\cal N} g_{0, k}^{(2)} x_{k, {\rm s}}^2\right)^2+{\kappa^2\over 4}}\right]x_{j, {\rm s}} &=& 0 \label{xss}.
\end{eqnarray}
In case all mechanical oscillators are located at local minima of the intracavity intensity, we have $g_{0, j}^{(2)}>0$ and the term in the square bracket in Eq.~(\ref{xss}) is always positive. In that case, the only solution has zero displacement for all the mechanical modes. However, if the mechanical oscillators are located at local maxima of the intracavity intensity, we have $g_{0, j}^{(2)}<0$ and for certain parameter regimes Eq.~(\ref{xss}) allows non-zero displacements.  For the case of identical coupling coefficients $g_{0, k}^{(2)}=g_{0}^{(2)}$ the non-zero displacements obey the equation
\begin{equation}
\sum_{k}^{\cal N}x_{k, {\rm s}}^2 = \frac{1}{|g_{0}^{(2)}|}\left[-\Delta_c\pm\sqrt{\frac{2|g_{0}^{(2)}||\eta|^2}{\omega_m}-\frac{\kappa^2}{4}}\right] \label{cdisplacement}.
\end{equation}
The stability of the steady-state positions can also be investigated via a qualitative analysis on the effective potential of the mechanics.  For quadratic interactions the effective potential obtained from Eq.~(\ref{cforce}) in the absence of dissipation is
\begin{eqnarray}
U_{\rm eff} &=&\frac{\hbar\omega_m}{2}\sum_{k}^{\cal N}x_{k}^2 \nonumber \\
&-&\frac{2\hbar|\eta|^2}{\kappa}\arctan\left[\frac{\Delta_c-\sum_{k}^{\cal N} g_{0, k}^{(2)} x_{k}^2}{\kappa/2}\right] \label{cpotential}.
\end{eqnarray}
Next we present three examples of the effective potential that highlight interesting features of the classical model with quadratic interactions, with a view to exploring the quantum version of the examples.

\begin{figure}[]
\includegraphics[width=0.48 \textwidth]{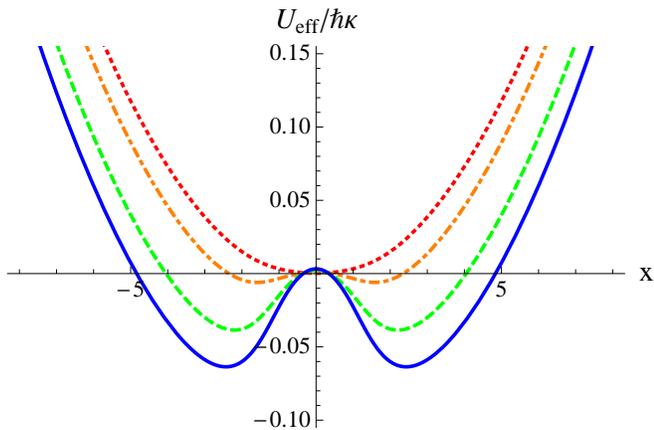}
\caption{\label{fig:single_potential}  (Color online) Effective potential $U_{\rm eff}(x)$, in units of $\hbar\kappa$, versus dimensionless position $x$ for a single mechanical oscillator quadratically coupled to the cavity mode for $|\eta|/\kappa = 0.05$ (red dotted line), $|\eta|/\kappa = 0.11$ (orange dot-dash line), $|\eta|/\kappa = 0.17$ (green dashed line), $|\eta|/\kappa = 0.20$ (blue solid line). Here $~\omega_m/\kappa = 0.01, ~g_{0, 1}^{(2)}/\kappa =-0.2,~\Delta_c/\kappa = -0.02$.}
\end{figure}
\begin{figure}
\includegraphics[width=0.48 \textwidth]{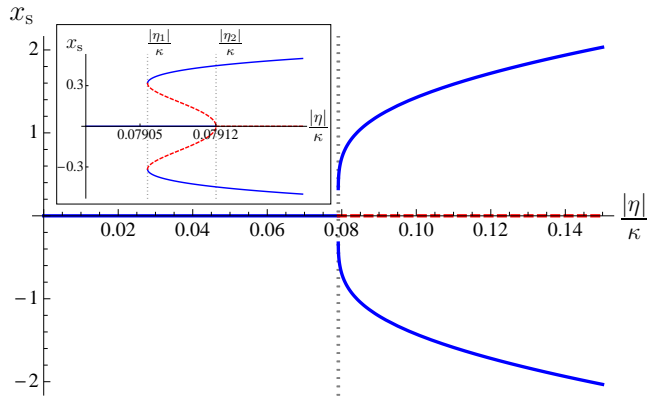}
\caption{\label{fig:single_displacement}  (Color online) Steady-state dimensionless positions $x_s$ for a single mechanical oscillator with quadratic coupling as a function of the normalized cavity pumping rate $|\eta|/\kappa$. The stable and unstable solutions are denoted by the solid blue and dashed red line, respectively.  Inset: zoomed in region around the bifurcation point. Here $~\omega_m/\kappa = 0.01, ~g_{0, 1}^{(2)}/\kappa =-0.2,~\Delta_c/\kappa = -0.02$.}
\end{figure}

For a first example we consider a single mechanical mode with quadratic coupling to the cavity mode.  The steady-state solutions $x_{k, {\rm s}}=x_s$  for this case are determined by Eq.~(\ref{cdisplacement}) with ${\cal N}=1$, which yields a fifth-order polynomial equation with up to five solutions ($x_s=0$ is always a solution).  In contrast to the case of linear coupling, the necessary condition for the mechanical oscillator to have multiple equilibrium position is simply that the single-photon coupling coefficient be negative, that is, that the mirror be trapped at a maximum of the intracavity intensity.  With a negative coupling coefficient the cavity pumping rate $|\eta|$ determines the number of physical solutions for $x_s$.  More specifically, below the first critical pumping rate
\begin{equation}
|\eta_1|=\sqrt{\frac{\omega_m\kappa^2}{8|g_{0, 1}^{(2)}|}} 
\end{equation}
the mechanical oscillator experiences a flattened harmonic-like effective potential and has only one stable equilibrium position at $x=0$. If the cavity pumping rate is increased such that
\begin{equation}
|\eta_1|<|\eta|<|\eta_2|,
\end{equation}
where the second critical pumping rate $|\eta_2|$ is given by
\begin{equation}
|\eta_2| = \sqrt{\frac{\omega_m}{2|g_{0, 1}^{(2)}|}\left[\Delta_c^2+\frac{\kappa^2}{4}\right]},
\end{equation}
the oscillator experiences a potential with five extrema, three stable equilibrium positions, including $x=0$, and two unstable equilibrium positions. For still stronger pumping rates larger than $|\eta_2|$, the zero displacement becomes unstable and the mechanical mode experiences a symmetric double well potential with only two stable equilibrium positions. Figure~\ref{fig:single_potential}  illustrates these features and shows the effective potential $U_{\rm eff}(x)$ versus dimensionless position $x$ for a single mechanical oscillator quadratically coupled to the cavity mode, at a maximum of the intracavity intensity. For our parameters the critical cavity pumping rates are $|\eta_1|/\kappa=0.07905$ and $|\eta_2|/\kappa=0.07912$.  

An alternative view of these results is shown in Fig.~\ref{fig:single_displacement} where we plot the allowed steady-state dimensionless positions $x_s$ as a function of the normalized cavity pumping rate for the  parameters of Fig.~\ref{fig:single_potential}.  The inset shows an expanded view of the plot around the critical cavity pumping rates.  What these results show is that around the critical pumping rates the system undergoes a subcritical bifurcation.  For larger cavity pumping rates the system displays two stable and energetically degenerate solutions. In any given realization of the system we expect that one or other of the two solutions will arise with equal probability if the system is initialized from noise. We explore the quantum dynamics in that regime in Sec.~\ref{ResultsQuad}.

For our second and third examples we consider two mechanical modes quadratically coupled to the cavity mode, either with both mechanical oscillators located at maxima of the intracavity intensity in which case the coupling constants are both negative
\begin{equation}
g_{0, 1}^{(2)} = g_{0, 2}^{(2)} \equiv -|g_{0}^{(2)}|  , 
\end{equation}
or with one oscillator located at a maximum and the other at a minimum, so that the coupling coefficients have opposite signs
\begin{equation}
g_{0, 1}^{(2)} = -g_{0, 2}^{(2)} \equiv |g_{0}^{(2)}|   .
\end{equation} 

\begin{figure}[]
\includegraphics[width=0.48 \textwidth]{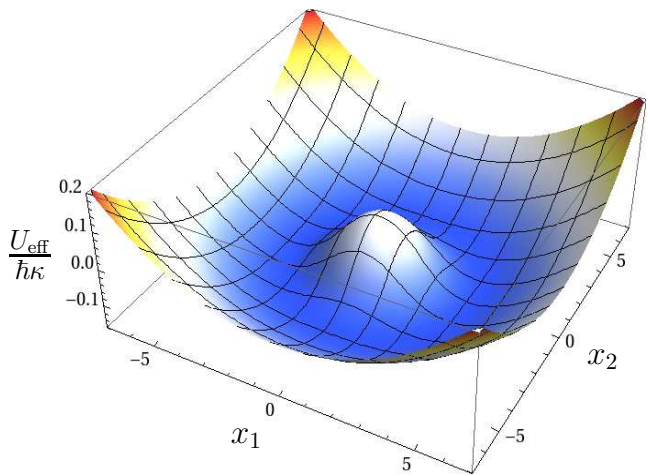}
\caption{\label{fig:potential_same_signs}  (Color online) Effective potential $U_{\rm eff}(x_1,x_2)$, in units of $\hbar\kappa$, versus the dimensionless positions $x_1$ and $x_2$ for equal and negative quadratic optomechanical coupling coefficients. Here $~\omega_m/\kappa = 0.01, ~g_{0}^{(2)}/\kappa =-0.2,~\Delta_c/\kappa = -0.02,~|\eta|/\kappa = 0.3$.}
\end{figure}

The effective potential $U_{\rm eff}(x_1,x_2)$ versus the dimensionless positions $x_1$ and $x_2$ of the mechanical oscillators can be obtained from Eq.~(\ref{cpotential}). An example of the first case, $g_{0, 1}^{(2)} = g_{0, 2}^{(2)}$, is shown in Fig.~\ref{fig:potential_same_signs}.  The key feature is that for large enough cavity pumping rates, the effective potential for the mechanical oscillators changes from a harmonic shape into a {\it sombrero} or Higgs potential, where the potential minimum is realized on a circle with radius
\begin{equation}
R = \sqrt{\frac{1}{|g_{0}^{(2)}|}\left[-\Delta_c\pm\sqrt{\frac{2|g_{0}^{(2)}||\eta|^2}{\omega_m}-\frac{\kappa^2}{4}}\right]}. 
\end{equation}

\begin{figure}[]
\includegraphics[width=0.48 \textwidth]{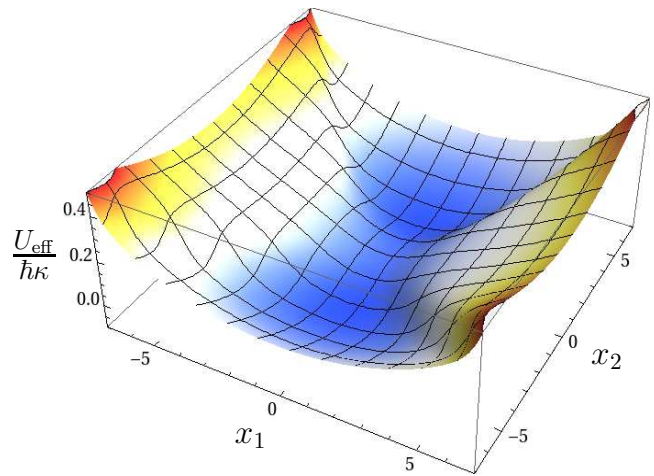}
\caption{\label{fig:potential_opposite_signs}  (Color online) Effective potential $U_{\rm eff}(x_1,x_2)$, in units of $\hbar\kappa$, versus the dimensionless positions $x_1$ and $x_2$ for quadratic interactions of opposite signs,  $~g_{0, 1}^{(2)}/\kappa = -g_{0, 2}^{(2)}/\kappa = 0.2, \omega_m/\kappa = 0.01, \Delta_c/\kappa = -0.02, |\eta|/\kappa = 0.22$. }
\end{figure}

Finally, Fig.~{\ref{fig:potential_opposite_signs}} illustrates the effective potential $U_{\rm eff}(x_1,x_2)$ for the case in which one of the mechanical oscillators is located at a local minimum and the other oscillator is located at a local maximum of the intracavity intensity, $g_{0, 1}^{(2)} = -g_{0, 2}^{(2)} $, with the cavity pumping rate chosen large enough to change the harmonic trapping potential to a double well potential for $x_2$ and to stiffen the harmonic trapping potential for $x_1$.   We shall explore quantum features of the two mode double well potential in Sec.~\ref{ResultsQuad}.

\section{Quantum effects}
We have so far investigated mean-field solutions of the mechanical modes, neglecting the effects of cavity field fluctuations on the mechanics. This analysis is valid when the cavity is sufficiently strongly pumped, and the cavity photon number is large enough, that the displacement of the mechanical oscillator is large compared to the natural harmonic oscillator length. However, this needs not be the case in the single photon strong coupling regime, where small photon numbers can still lead to optomechanical couplings comparable to the mechanical frequency, and both cavity and mechanical systems can be in the deep quantum regime.   In this case quantum fluctuations become significant and can enter the quantum dynamics of the mechanics in interesting ways.

The remainder of this article presents numerical results that illustrate the dynamics of one or two mechanical modes coupled to the single cavity mode via either linear or quadratic optomechanical interactions in the single-photon strong-coupling regime  $g_{0, j } \ge \omega_m > \gamma$. We concentrate on the prevalent case where the cavity decay rate $\kappa$ is much larger than the single-photon coupling coefficient, so that $\kappa \gg g_{0, j} \ge \omega_m \gg \gamma$.  In this regime, the cavity field follows the dynamics of the mechanical mode and nonlinear quantum effects can be observed at a single-photon level. This regime can be realized in optomechanical systems involving ultracold atoms \cite{atom1, atom2, atom3, atom4} but has not yet been reached in current state-of-the-art micromechanical systems. For concreteness in our quantum simulations we adopt representative parameters from the experiment in Ref.~\cite{atom2}, namely, the cavity decay rate $\kappa= 2\pi\times2.6~\mathrm{MHz}$, mechanical frequency $\omega_m= 2\pi\times15.2~\mathrm{kHz}$, and effective single-photon coupling coefficient $g_{0, 1}= 2\pi\times0.5~\mathrm{MHz}$. In units such that $\kappa=1$, these are equivalent to $\omega_m= 0.6\times10^{-2}$, $g_{0, 1}= 1.9\times10^{-1}$. We concentrate on the case where the laser is red-detuned with respect to the cavity resonance.

Before presenting results based on a direct numerical integration of the master equation for the oscillator-light system -- that is, without adiabatic elimination of the optical field -- we discuss briefly the Heisenberg-Langevin equations of motion for the mechanical oscillators in order to capture more intuitively perhaps the effects of cavity fluctuations on the dynamics of the mechanical mode.  

\subsection{Heisenberg-Langevin equations}
Starting from the quantum mechanical Heisenberg-Langevin equations for the cavity and mechanical modes of Sec.~\ref{Basic}, in the regime where the cavity decay is the dominant rate we may adiabatically eliminate the cavity mode while retaining the quantum noise terms.  For the case of linear coupling this yields the effective Heisenberg-Langevin equations of motion for the mechanical oscillator
\begin{eqnarray}
\dot{\hat{x}}_j &=& \omega_m\hat{p}_j, \\
\dot{\hat{p}}_j &=& -\omega_m\hat{x}_j +\frac{g_{0, j}|\eta|^2}{\left(\Delta_c+\sum_{k}^{\cal N}g_{0, k}\hat{x}_k\right)^2+\kappa^2/4} \nonumber\\
&+&\left[\frac{g_{0, j}\eta^{*}\hat{\zeta}}{i\left(\Delta_c+\sum_{k}^{\cal N}g_{0, k}\hat{x}_k\right)+\kappa/2}+H.c.\right] \nonumber\\
&+&g_{0, j}\hat{\zeta}^{\dag}\hat{\zeta}-\frac{\gamma}{2}\hat{p}_j+\hat{\xi} 
\label{linear_qforce}, 
\end{eqnarray}
where the cavity noise operator $\hat{\zeta}$ involving the cavity input noise $\hat{a}_{\rm in}$ is 
\begin{equation}
\hat{\zeta}(t) \approx \sqrt{\kappa}\int_{0}^{t}d\tau e^{(i\Delta_c-\kappa/2)(t-\tau)}\hat{a}_{\rm in}(\tau). 
\end{equation}
Following the same procedure for the case of quadratic coupling yields the effective Heisenberg-Langevin equations
\begin{eqnarray}
\dot{\hat{x}}_j &=& \omega_m\hat{p}_j, \\
\dot{\hat{p}}_j &=& -\omega_m\hat{x}_j -\frac{2g_{0, j}^{(2)}|\eta|^2}{\left(\Delta_c-\sum_{k}^{\cal N}g_{0, k}^{(2)}\hat{x}_k^2\right)^2+\kappa^2/4}\hat{x}_j \nonumber\\
&-&\left[\frac{2g_{0, j}^{(2)}\eta^{*}\hat{\zeta}}{i\left(\Delta_c-\sum_{k}^{\cal N}g_{0, k}^{(2)}\hat{x}_k^2\right)+\kappa/2}+H.c.\right]\hat{x}_j \nonumber\\
&-&2g_{0, j}^{(2)}\hat{\zeta}^{\dag}\hat{\zeta}\hat{x}_j-\frac{\gamma}{2}\hat{p}_j+\hat{\xi} .
\label{qforce}
\end{eqnarray}
For both linear and quadratic interactions the second term on the right-hand-side of the equations of motion~(\ref{linear_qforce}) and~(\ref{qforce}) for the mechanical momenta is independent of the cavity noise, but rather derives from the effective potential.  In contrast, the third and fourth terms explicitly involve the additional random forces due to the quantum fluctuations of the optical field. These are in addition to the intrinsic random forces associated with their direct coupling to a heat bath. Importantly, in both the case of linear and quadratic coupling the additional noise experienced by the mechanical modes is multiplicative, with consequences that will be discussed in the next section. 

\subsection{Master equation}
The master equation describing the evolution of the total density operator prior to adiabatic elimination of the cavity field is~\cite{textbook} 
\begin{equation}
\frac{d}{dt}\hat{\rho} = -\frac{i}{\hbar}[\hat{H}, \hat{\rho}] + \frac{\kappa}{2}{\cal D}[\hat{a}]\hat{\rho} +\frac{\gamma}{2}\sum_{k=1}^{\cal N}{\cal D}[\hat{b}_k]\hat{\rho},
\end{equation}
where $\hat{b}_k$ is the annihilation operator for the $k$-th mechanical mode, and we have assumed that the cavity and mechanical modes are both coupled to reservoirs at zero temperature for simplicity, so that 
\begin{equation}
{\cal D}[\hat{o}]\rho = (2\hat{o}\rho\hat{o}^{\dag}-\hat{o}^{\dag}\hat{o}\rho-\rho\hat{o}^{\dag}\hat{o}).
\end{equation}
For small enough numbers of photons and phonons and small number of mechanical modes the size of the relevant Hilbert space remains manageably small and it is possible to solve that master equation directly by brute force, without resort to the adiabatic elimination of the cavity field. We proceed by expanding the density matrix in the Fock states basis $\{n_a, n_{b_1},.., n_{b_{\cal N}}\}$ as
\begin{equation}
\hat{\rho} = \sum\rho_{n_a, m_a, n_{b_1}, m_{b_1},..}|n_a, n_{b_1},..,n_{b_{\cal N}}\rangle\langle m_a, m_{b_1},..,m_{b_{\cal N}}|,
\nonumber
\end{equation}
and verify that the Hilbert space is large enough to avoid boundary issues and that the norm of the density operator is preserved at all times. 

\section{Results}\label{Results}

\subsection{Linear interactions}\label{ResultsLin}

\subsubsection{Single-mode mechanics}

\begin{figure}[]
\includegraphics[width=0.48 \textwidth]{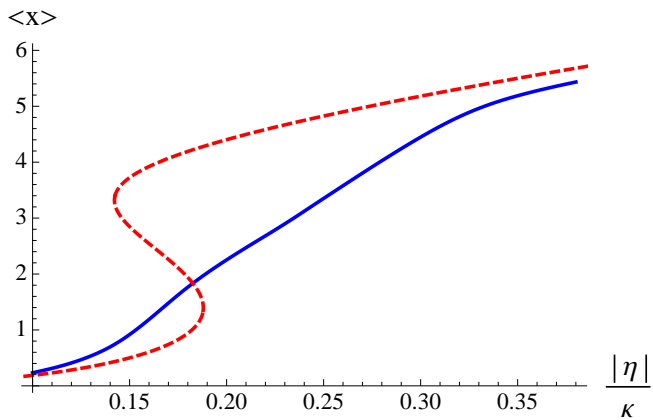}
\caption{\label{fig:No_bistability}   (Color online) Expectation value $\langle \hat x \rangle$ versus normalized cavity pumping rate $|\eta|/\kappa$ (solid blue line) for a single mechanical oscillator and linear optomechanical interaction. Here $~\omega_m/\kappa = 0.01, ~g_{0, 1}/\kappa =0.3,~\Delta_c/\kappa = -1.5,~\gamma/\kappa = 0.002$. The red dashed curve shows the corresponding classical bistable solution.}
\end{figure}

This subsection considers the case of a single mechanical mode linearly coupled to the optical field, using the same parameters as in Fig.~\ref{fig:linear_single_potential}, and addresses how quantum fluctuations, in particular the multiplicative noise of Eq.~(\ref{linear_qforce}), impact the mean-field bistable behavior. It is known that deep in the quantum regime quantum fluctuations eradicate the possibility that the system dwells in one or other of the two classically allowed states~\cite{QED_bistability1}. This was demonstrated experimentally at the single atom and single photon level in cavity QED experiments with ultracold atomic beams~\cite{QED_bistability2,footnote0}.  Not surprisingly, a similar situation occurs here for the center-of-mass of the mechanics. This is illustrated in Fig.~\ref{fig:No_bistability} which shows the quantum expectation value $\langle x \rangle$ of the dimensionless position operator for the mechanical oscillator in steady state versus the normalized cavity pumping rate $|\eta|/\kappa$ (solid blue line), along with the classically allowed positions (red dash line). Recall that the negative slope region of the classical solution is unstable.  
\begin{figure}[]
\includegraphics[width=0.48 \textwidth]{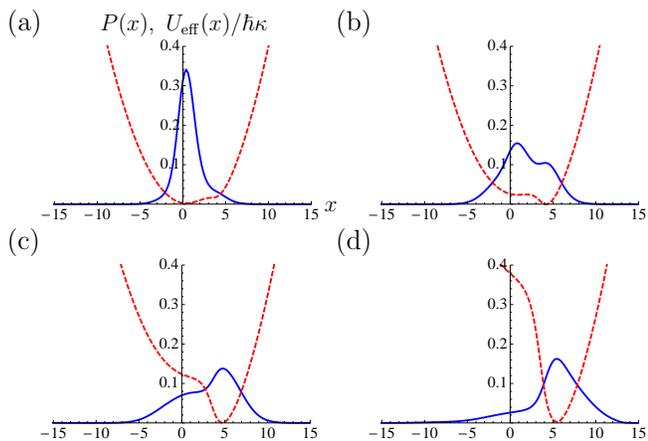}
\caption{\label{fig:Linear_singlemode}    (Color online) Steady-state position probability distribution $P(x)$ versus dimensionless position $x$ (solid blue lines) of the mechanical oscillator along with the effective potential $U_{\rm eff}(x)$ (red dash lines), in units of $\hbar\kappa$, for the normalized cavity pumping rates $|\eta|/\kappa$ (a) $0.14$, (b) $0.18$, (c) $0.24$, (d) $0.34$ for a single mechanical oscillator and linear optomechanical interaction. Here $~\omega_m/\kappa = 0.01, ~g_{0, 1}/\kappa =0.3,~\Delta_c/\kappa = -1.5$, and $\gamma/\kappa = 0.002$.}
\end{figure} 

To further clarify the washing out of the mechanical bistability the solid blue lines in Fig.~\ref{fig:Linear_singlemode} show the steady-state position probability distribution 
\begin{equation}
P(x) \equiv P(x, t \rightarrow \infty) = \langle x|\hat{\rho}_{m}(t \rightarrow \infty)|x \rangle,
\end{equation}
where $\hat{\rho}_{m}(t \rightarrow \infty)$ is the reduced density matrix for the mechanical subsystem in the steady state and $|x\rangle$ is the eigenstate of the dimensionless position operator $\hat{x}$, for several values of the normalized cavity pumping rate $|\eta|/\kappa$, the dashed red lines being the corresponding effective classical potential $U_{\rm eff}(x)$ in Eq.~(\ref{linear_cpotential}). As expected, Fig.~{\ref{fig:Linear_singlemode}}(b) displays a bimodal probability density for a cavity pumping rate for which the classical theory predicts bistability.  We note, however, that the absolute peak of the $P(x)$ distribution does not correspond to the absolute minimum of the classical potential, as would be expected on the basis of additive noise.  This can be intuited by realizing that, for a single mechanical mode, the cavity noise operator $\hat\zeta$ in the third term of Eq.~(\ref{linear_qforce}) appears in conjunction with a cavity resonant denominator involving the mode position operator, meaning that this noise source is multiplicative.  The (classical) lower branch of the bistability curve therefore corresponds to lower intracavity fields than the upper branch, and therefore less quantum noise. For this reason, the shallower minimum of the potential is rendered more stable than the deeper minimum against quantum noise. As such, this behavior is a direct consequence of the multiplicative nature of the noise. 

\subsubsection{Two-mode mechanics}
\begin{figure}[]
\includegraphics[width=0.48 \textwidth]{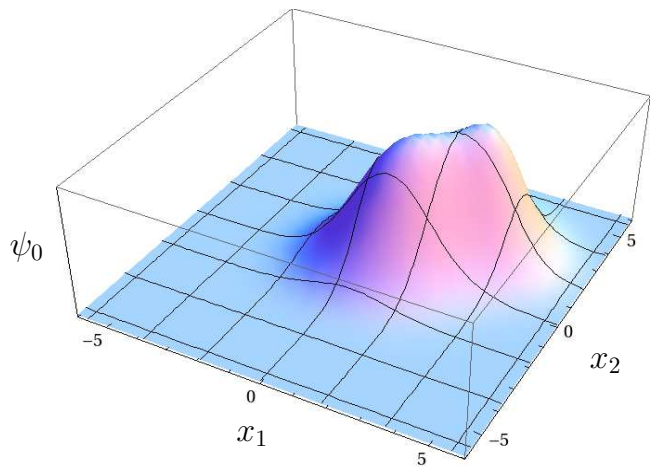}
\caption{\label{fig:Linear_two_ground}  (Color online) Ground state wave function $\psi_0(x_1,x_2)$ as a function of the dimensionless positions $x_1$ and $x_2$. Same parameters as in Fig.~\ref{fig:lin_two_pot}. }
\end{figure}

We now turn to the case of two mechanical modes of equal frequency $\omega_m$ and  equal linear optomechanical coupling to the optical field mode.   To set the stage we first ignore cavity field fluctuations and determine the quantum mechanical ground state wave function $\psi_0(x_1,x_2)$ of the effective potential $U_{\rm eff}(x_1, x_2)$, given by
\begin{equation}
-\frac{\hbar\omega_m}{2}\left[\frac{d^2}{dx_1^2}+\frac{d^2}{dx_2^2}\right]\psi_0 +U_{\rm eff}(x_1, x_2)\psi_0 = E_0\psi_0,
\end{equation}
where $E_0$ is the energy eigenvalue, using the imaginary time propagation method.  The ground state wave function, assumed real and positive, is plotted in Fig. \ref{fig:Linear_two_ground}. As expected from the effective potential, it has two peaks localized at the local minima of $U_{\rm eff}(x_1, x_2)$. 
\begin{figure}[]
\includegraphics[width=0.48 \textwidth]{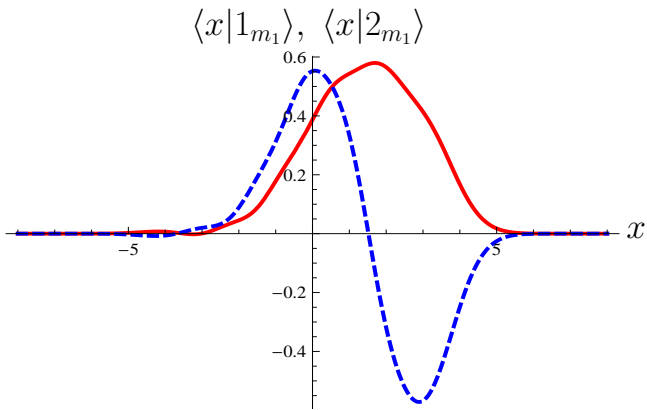}
\caption{\label{fig:Linear_schmidt_basis}  (Color online) Position representation of the Schmidt basis states $\langle x|1_{m_1}\rangle$ (red solid line) and $\langle x|2_{m_1}\rangle$ (blue dashed line) for the two mechanical  oscillators in the ground state $\psi_0(x_1, x_2)$. Same parameters as in Fig.~\ref{fig:lin_two_pot}.}
\end{figure}

We next make use of the Schmidt decomposition of $\psi_0(x_1, x_2$) in order to determine whether the two mechanical oscillators in the ground state can be separated or not. It is known that subsystems are entangled if their Schmidt number, or the number of nonzero Schmidt coefficients, is greater than unity~\cite{Schmidt_number}. Decomposing $\psi_0(x_1, x_2)$  as
\begin{equation}
|\psi_0\rangle = \sum_i\lambda_i |i_{m_1}\rangle |i_{m_2}\rangle,
\end{equation}
where $\lambda_i$ are the Schmidt coefficients with respect to the basis $|i_{m_1}\rangle |i_{m_2}\rangle$, $m_1$ and $m_2$ labeling the two mechanical oscillators, we find for the case at hand that the non-zero Schmidt coefficients $\lambda_1=0.96,~\lambda_2=0.29,~\lambda_3=0.02$, and $~\lambda_4=0.02$ in descending order.   Since the Schmidt number is greater than unity the ground state of the two oscillators is entangled, with the state dominated by the first two Schmidt states $|1_{m_1}\rangle |1_{m_2}\rangle$ and $|2_{m_1}\rangle |2_{m_2}\rangle$. Their position representation is illustrated in Fig.~\ref{fig:Linear_schmidt_basis}. 

\begin{figure}[]
\includegraphics[width=0.48 \textwidth]{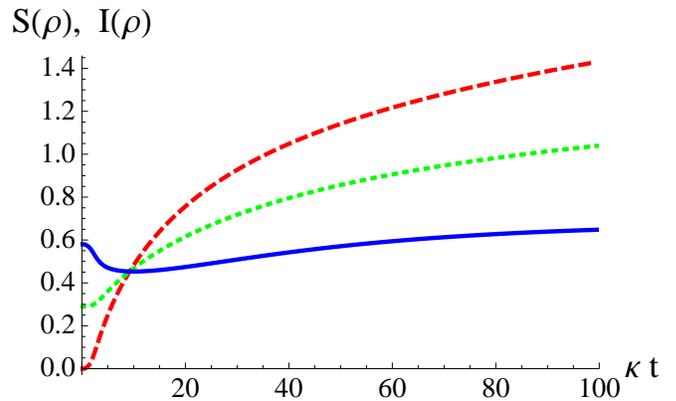}
\caption{\label{fig:Linear_entropy} 
 (Color online) Von Neumann entropies and quantum mutual information of the mechanical system as a function of normalized time $\kappa t$. Green dotted line: entropies of the individual mechanical oscillators; red dashed line:  their quantum joint entropy; blue solid line: quantum mutual information of the composite system. Here $\gamma_1=\gamma_2=2\times 10^{-3}\kappa$,  other parameters as in Fig.~\ref{fig:lin_two_pot}.}
\end{figure}

So far we have neglected the effects of cavity mode fluctuations. In order to ascertain the contribution of cavity fluctuations and decoherence on the dynamics of the mechanical system, we finally determine the evolution of the initial state $\psi_0(x_1,x_2)$ including cavity and mechanical damping, and quantify the correlations between the mechanical oscillators through their quantum mutual information
\begin{equation}
I(\hat{\rho}_m) = S(\hat{\rho}_{m_1})+S(\hat{\rho}_{m_2})-S(\hat{\rho}_{m}).
\end{equation}
Here $S(\hat{\rho}_{m})$ is the quantum joint entropy -- or simply entropy -- of the composite mechanical system, and 
$S(\hat{\rho}_{m_1})$ and $S(\hat{\rho}_{m_2})$ are the von Neumann entropies of the individual mechanical oscillators, with
\begin{equation}
S(\hat{\rho}) = -{\rm Tr}[\hat{\rho}\ln\hat{\rho}]. 
\end{equation}
Figure~\ref{fig:Linear_entropy} shows the  time evolution of the entropy of the individual mechanical oscillators (green dotted line), which is the same for both oscillators, their joint quantum entropy (red dashed lines), and their quantum mutual information (blue solid line). The initial ground-state entanglement between the two mechanical oscillators is apparent from the fact that the quantum entropies of the mechanical subsystems are nonzero while their joint entropy vanishes~\cite{Schmidt_number}. Both the joint entropy and the entropy of the individual oscillators tend to increase with time due to both the random radiation pressure variations arising from cavity intensity fluctuations and mechanical damping. We also note that the quantum mutual information between the two oscillators (blue solid line) is maintained in the long-time limit, indicative of the correlations between the two mechanical subsystems resulting from their optically mediated interaction via a common cavity mode.

\subsection{Quadratic interactions}\label{ResultsQuad}

\subsubsection{Single-mode mechanics}
As in Sec.~\ref{classical} we first consider a single mechanical mode with negative single-photon optomechanical coupling coefficient. We determined (see Fig.~\ref{fig:single_potential}) that classically the system undergoes a subcritical bifurcation for sufficiently large cavity pumping rate.  Here we investigate the impact of quantum fluctuations on the associated dynamics.

\begin{figure}[]
\includegraphics[width=0.48 \textwidth]{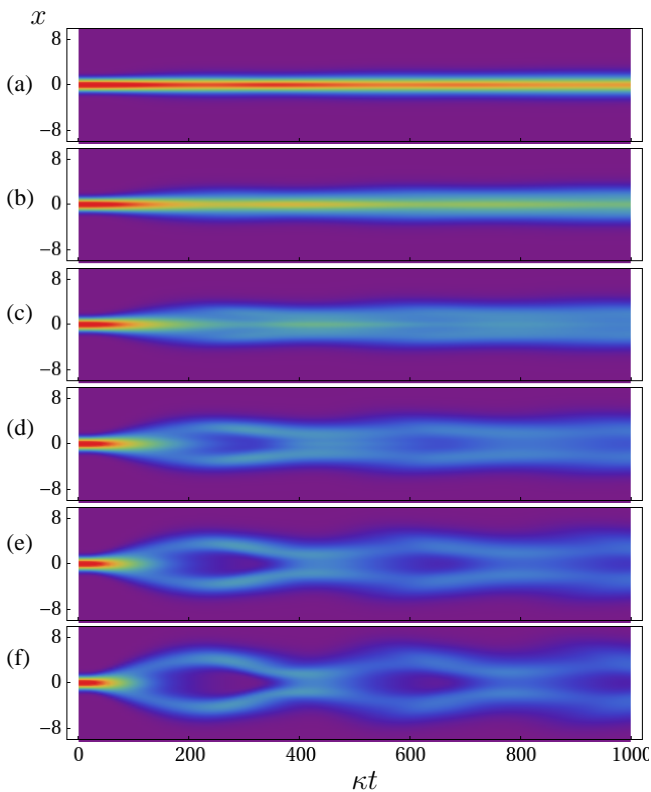}
\caption{\label{fig:ground_phase}   (Color online) Normalized time $(\kappa t)$ evolution of the spatial probability $P(x,t)$ for a mechanical mode initially in its ground state, and for the optical pumping rates $|\eta|/\kappa$ (a) $0.05$, (b) $0.08$, (c) $0.11$, (d) $0.14$, (e) $0.17$, and (f) $0.20$.  In each panel the vertical axis is the dimensionless position $x$, and $P(x,t)$ is color coded.  See the potential $U_{\rm eff}$ of Fig.~\ref{fig:single_potential} and Fig.~\ref{fig:single_displacement}, which are for the same set of parameters, for reference.  Here $\gamma = 10^{-3}\kappa$.}
\end{figure}

Figure~\ref{fig:ground_phase} shows the normalized time $(\kappa t)$ evolution of the oscillator spatial probability distribution $P(x, t)$ for a variety of cavity pumping rates, obtained by direct numerical solution of the master equation for the oscillator-field system.  In this example the mechanical mode is initially in its ground state and the optical field in the vacuum, and the cavity pumping rate is switched on suddenly at $t=0$ to a constant value $\eta$. The successive panels  show the effect of increased $|\eta|/\kappa$. For pumping below the bifurcation point -- which occurs at $|\eta|/\kappa \simeq 0.08$ for the parameters of the figure -- see panels (a) and (b), the variance in position of the mechanics increases as a consequence of the flattening of the effective potential as the bifurcation point is approached from below, see Fig.~\ref{fig:single_potential}.  

For cavity pumping rates past the bifurcation point, see panels (c)-(f), $U_{\rm eff}$ is a double-well potential with the zero displacement point $x=0$ unstable and two stable and degenerate minima.  From a classical perspective, above that point we expect fluctuations to drive the system into one or other of these two minima. Panels (d)-(f) of Fig.~\ref{fig:ground_phase} show that quantum mechanically the mechanical mode, initially localized around $x=0$, undergoes oscillations involving both potential minima.  Taking panel (f) as an example, we see that the initial quantum wave packet splits symmetrically between both wells, reverses at the turning point of the double-well potential at $\kappa t\approx 200$, and the split wave-packet components recombine at $\kappa t\approx 400$.  Bifurcation-induced wave-packet splitting and subsequent recombination requires the cavity pumping rate to be sufficiently above the bifurcation point so that the split wave-packet components become well separated spatially: This conforms to the usual notion that a quantum phase transition will be smoothed out close to the bifurcation point, in comparison to a classical bifurcation that occurs discretely. This basic process can repeat several times but with diminishing contrast due to the combined action of optical and mechanical decoherence, see panels (e)-(f). 
\begin{figure}[]
\includegraphics[width=0.48 \textwidth]{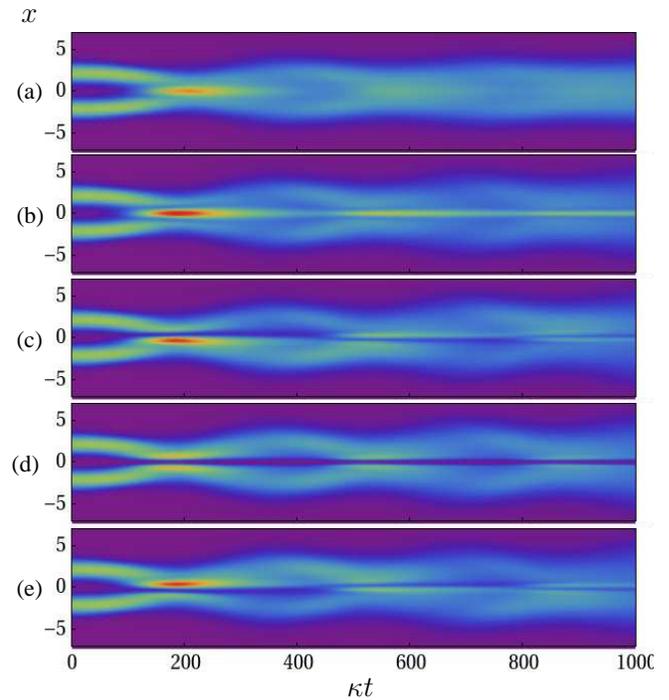}
\caption{\label{fig:cat}  (Color online) Normalized time $(\kappa t)$ evolution of the position probability distribution $P(x, t)$ for an oscillator initially in the cat state (\ref{cat}) with $\beta_0 = 1.5$ and the relative phases (a) $\phi_0 = 0$,  (b) $\phi_0=0$, (c) $\phi_0 =\pi/2$, (d) $\phi_0 =\pi$, and (e) $\phi_0 =3\pi/2$. In each panel the vertical axis is the dimensionless position $x$, and $P(x,t)$ is color coded.  Same parameters as in Fig. ~\ref{fig:single_potential}, with cavity pumping rate $|\eta|/\kappa = 0.17$. The mechanical decay rate is $\gamma= 10^{-3} \kappa$ in panel (a) and $\gamma=10^{-6} \kappa$ in panels (b)-(e).}
\end{figure}

The wave-packet splitting and subsequent recombination shown between $\kappa t=[0,400]$ in panel (f) of Fig.~\ref{fig:ground_phase} is reminiscent of a Mach-Zehnder interferometer. This suggests exploiting this system for observing quantum interferences, provided that decoherence remains manageable.  To explore the relative contributions of the optical and mechanical damping to decoherence we now calculate the evolution of the mechanical mode initially prepared in a coherent superposition of wave packets localized at non-zero displacements $\pm \beta_0$  with different relative phase $\phi_0$
\begin{equation}
\psi_m(0) = \frac{1}{\sqrt{2}}(|\beta_0\rangle+ e^{i\phi_0}|-\beta_0\rangle),
\label{cat}
\end{equation}
and for two different values of the mechanical damping. The idea of this simulation is that the two components of the coherent superposition are representative of the wave-packet components resulting from the bifurcation-induced splitting at $\kappa t\simeq 200$ in panel (f) of Fig.~\ref{fig:ground_phase}.  The subsequent time evolution of the probability density $P(x, t)$ for these ``cat states'' is plotted in Fig.~\ref{fig:cat} for a cavity pumping rate beyond the bifurcation point. Panel (a) is for $\gamma = 10^{-3} \kappa$ and $\phi_0=0$, while the subsequent panels are for $\gamma = 10^{-6} \kappa$ and various values of $\phi_0$. 

The bottom four panels show that for the case of negligible mechanical damping (on the time scale of the plots) the probability near zero displacement depends on the initial relative phase $\phi_0$, a clear signature of a quantum interference effect.  Interferences reappear periodically for longer times, but with a slowly decreasing amplitude due to the decoherence resulting from the quantum fluctuations of the optical field.  What is perhaps surprising is however that the interferences subsist for remarkably long times, thousands of cavity decay times $\kappa^{-1}$. That quantum coherence can persist on such long time scales is attributed to the coherent pumping of the cavity mode. It is known, see e.g. Refs.~\cite{Meystre90,Slosser90}, that the coherent pumping of Schr{\"o}dinger cats can result in maintaining their coherence for arbitrarily long times. In the specific case of coherently driven micromasers for example, it was shown that the onset of these superpositions resembles a second-order phase transition, with the control parameter being the ratio of the atomic injection rate to the cavity damping rate. In contrast, panels (a) and (b) illustrate the effect intuitively expected from mechanical dissipation. All parameters are identical in these panels, except that $\gamma = 10^{-3}\kappa$ in (a) and $\gamma=10^{-6}\kappa$ in (b). As expected, the first interference peak visible in panel (b) is already significantly reduced in case (a) after a time of about $0.2 \gamma^{-1}$, and all but extinguished after a time $\gamma^{-1}$.  

\begin{figure}[]
\includegraphics[width=0.48 \textwidth]{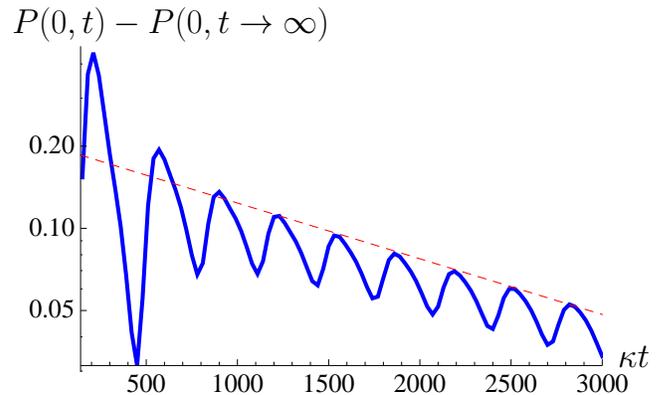}
\caption{\label{fig:center}  (Color online) Semi-log plot of the position probability distribution $P(x=0, t)-P(x=0,t \rightarrow \infty)$ as a function of normalized time $(\kappa t)$ (blue, solid curve). The red straight line is a fit to the long time maxima of the distribution. Same parameters as in Fig.~\ref{fig:cat}(b).}
\end{figure}

Returning to the remarkably slow optically induced decoherence, Fig.~\ref{fig:center} shows on a semi-log scale $P(x=0,t) - P(x=0,t \rightarrow \infty)$ versus normalized time $\kappa t$ for a case of negligible mechanical damping, $\gamma=10^{-6}\kappa$, the red dashed straight line being a fit through the peak maxima that illustrates an effective exponential decay rate about 3 order of magnitude slower than $\kappa^{-1}$ for $\kappa t > 1000$, but that starts off faster and non-exponentially~\cite{footnote}. 

We attribute the initial decay to the multiplicative nature of the noise due to cavity mode fluctuations appearing in the third term on the right-hand-side of Eq.~(\ref{qforce}).  First, we observe that this term gives rise to fluctuations in the frequency experienced by the mechanical mode, and it is known that such frequency fluctuations can translate into an effective decay~\cite{textbook}.  Second, it has a resonant denominator that assumes its smallest value, and hence gives the largest loss, when the mechanical mode has a position in the vicinity of the minima of the double-well effective potential, whereas the loss will be relatively small when the oscillator is in the vicinity of $x=0$.  With reference to Fig.~\ref{fig:cat}(b) we see that between the first and second peaks, that occur at $\kappa t \approx 190$ and $\kappa t \approx 540$, the probability density $P(x,t)$ undergoes a transient and recurs close to the initial form $P(x,0)$, which is centered around the minima of the effective potential. Thus there is sizable loss between the first two peaks.  However, between subsequent pairs of neighboring peaks there is less of a recurrence, and the decay rate between peaks decreases with increasing time.  Eventually $P(x,t)$ approaches a near steady-state and the loss rate becomes exponential.

We remark that optical decoherence can also be reduced by using a pulsed rather than a continuous wave laser to excite the cavity, pulsed optomechanics having been previously studied in the context of squeezing of the position uncertainty of a mechanical oscillator \cite{pulse}. Thus, using the bifurcation-induced wave-packet splitting followed by evolution of the subsequent wave packet in the harmonic potential of the mechanical mode offers a route to observing quantum interference effects alluded to here over times much longer than the optical decoherence time  -- but of course shorter than the inverse mechanical decay rate.  This suggests that a single mode with quadratic optomechanical coupling, coherent driving and extremely slow mechanical damping is a viable candidate for producing quantum interference effects resulting from the bifurcation induced splitting. 

\subsubsection{Two-mode mechanics}

\paragraph{Equal Couplings}
We finally turn to the case of two mechanical modes with quadratic interactions, considering as in section \ref{classical} both the cases of equal negative coupling coefficients and of coupling coefficients of equal magnitude but opposite sign. In the first case the classical dynamics of the mechanics is captured by the effective potential $U_{\rm eff}(x_1,x_2)$ of  Fig.~\ref{fig:potential_same_signs}, which has the form of a sombrero (or Higgs) potential for sufficiently large cavity pumping rate.  To exploit its rotational symmetry we introduce the angular momentum in the $(x_1,x_2)$ plane
\begin{equation}
\hat{L}_{\phi} = \hat{x}_1\hat{p}_2-\hat{x}_2\hat{p}_1,
\end{equation}
with Heisenberg-Langevin equation of motion 
\begin{equation}
\dot{\hat{L}}_{\phi} = -\frac{\gamma}{2}\hat{L}_{\phi}+\hat{\xi}_{\phi},
\label{L Heisenberg}
\end{equation}
and we have introduced the noise operator 
\begin{equation}
\hat{\xi}_{\phi} = \hat{x}_1\hat{\xi}_2-\hat{x}_2\hat{\xi}_1.
\end{equation}
As expected from the symmetry of the potential,  $\hat{L}_{\phi}$ is a constant of motion in the absence of mechanical dissipation. Importantly, the angular momentum is insensitive to  cavity fluctuations and the associated decoherence, assuming as we have done that both mechanical modes are subject to the same optomechanical coupling. 

From Eq.~(\ref{L Heisenberg}) we have
\begin{equation}
\langle\hat{L}_{\phi}\rangle (t) = e^{-\gamma t/2}\langle\hat{L}_{\phi}\rangle (0).
\end{equation}
showing that the mean angular momentum decays to zero due to mechanical phase diffusion with the characteristic time scale of $\gamma^{-1}$. 

\paragraph{Opposite Couplings}

\begin{figure}[]
\includegraphics[width=0.48 \textwidth]{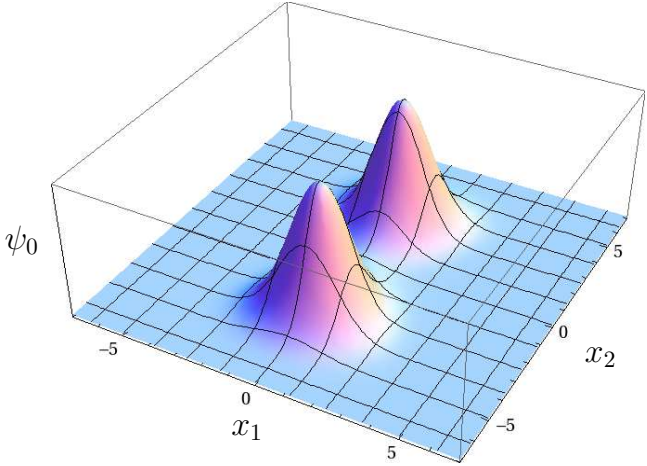}
\caption{\label{fig:opposite_signs_ground_state}   (Color online) Ground state of the effective potential describing the motion of the two mechanical modes for the case of quadratic interactions and coupling coefficients of equal magnitude but opposite sign.  Same parameters as in Fig.~\ref{fig:potential_opposite_signs}. }
\end{figure}
Perhaps more interesting is the situation where the two modes have optomechanical coupling constants of equal magnitude but opposite signs and are governed classically by the effective potential of Fig.~{\ref{fig:potential_opposite_signs}}.  The corresponding quantum mechanical ground state is plotted in Fig.~\ref{fig:opposite_signs_ground_state}. Not surprisingly, it is symmetric about both the lines $x_1=0$ and $x_2=0$,  and exhibits two peaks localized at the local minima of $U_{\rm eff}$. Physically this state corresponds to the oscillator  ``2''  being in a ``cat state'', whereas the oscillator ``1'' is a Gaussian centered at the origin $x_1=0$.  The Schmidt number for this ground state is found numerically to be equal to one, which implies that it is separable.

\begin{figure}[]
\includegraphics[width=0.48 \textwidth]{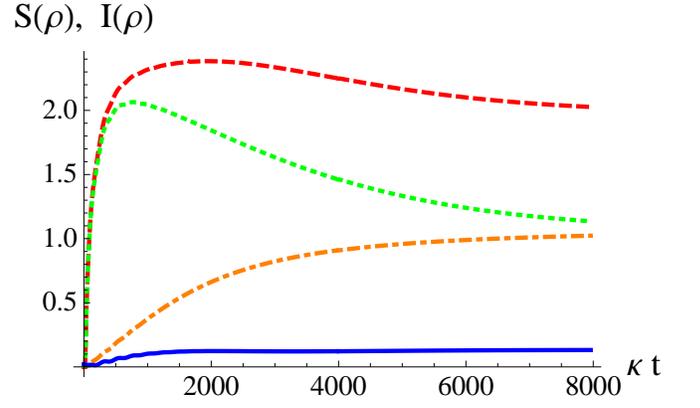}
\caption{\label{fig:Quadratic_entropy}  (Color online) Von Neumann entropies of the mechanical oscillators ``1'' (orange dot-dash line) and  ``2'' (green dotted line), their joint quantum entropy (red dash line), and mutual quantum information (blue solid line).  Same parameters as in Fig.~\ref{fig:potential_opposite_signs} with $\gamma_1/\kappa=\gamma_2/\kappa=10^{-3}$.}
\end{figure}

Figure~{\ref{fig:Quadratic_entropy}} shows the time evolution of the von Neumann entropies of oscillators ``1'' (orange dot-dashed line) and ``2'' (green dotted line), their joint  entropy (red dashed line), and mutual quantum information (blue solid line).  All entropies being initially equal to zero confirms that the ground state is separable, with both subsystems in pure states~\cite{Schmidt_number}.  Under the influence of quantum noise from both the optical field and the mechanics the entropy of the oscillators then increases, with oscillator ``1''  experiencing an entropy increase that is much slower than oscillator ``2''., This is not surprising, since due to its cat-like nature the second oscillator is expected to be much more sensitive to decoherence. Eventually the increase in entropy of oscillator ``2''  reverses, and asymptotically the mechanics reaches a situation where entropy is distributed almost equally between the two oscillators. As was the case for linear optomechanical coupling, see Fig.~\ref{fig:Linear_entropy}, the growth of their quantum mutual information (blue solid line) with time shows that the cavity fluctuations in fact correlate the two initially uncorrelated mechanical oscillators, that is, dissipation builds correlation via interaction of the two oscillators with a common light field or bath. 

\begin{figure}[]
\includegraphics[width=0.48 \textwidth]{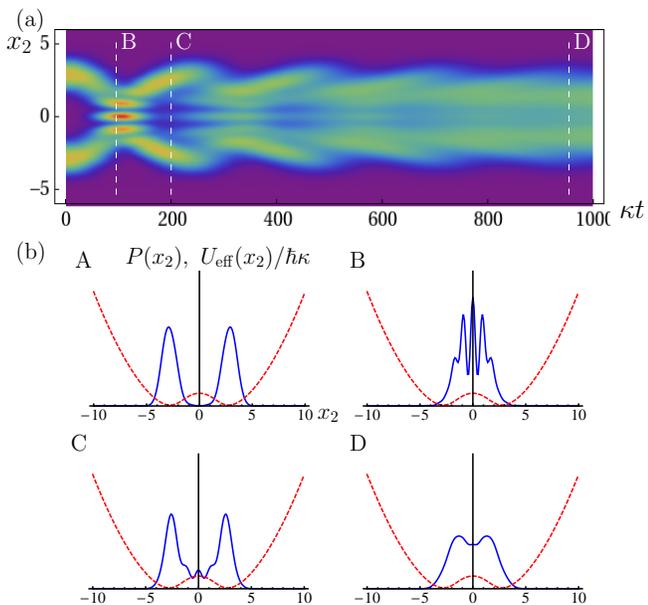}
\caption{\label{fig:p2}  (Color online) (a) Normalized time $(\kappa t)$ evolution of the marginal probability distribution $P(x_2, t)$ where the vertical axis is the dimensionless position $x$, and $P(x,t)$ is color coded, and (b) the effective potential (red line), in units of $\hbar\kappa$, and probability distribution $P(x_2, t)$ (blue line) for the times indicated in panel (a), with A corresponding to the initial time $t=0$. Same parameters as in Fig.~\ref{fig:Quadratic_entropy}, $\gamma_1/\kappa=\gamma_2/\kappa=10^{-3}$. }
\end{figure}

The absence of mutual coherence between the two oscillators in the ground state begs the question of the extent to which the spatial coherence of oscillator 2 -- which is initially in a cat-like state associated with the double-well effective potential along axis $x_2$ -- survives in the presence of decoherence.  To explore this we computed the time dependence of the marginal probability density $P(x_2,t)$, see Fig.~{\ref{fig:p2}}.  As expected the initial interference fringes wash out in the long term limit, but similarly to the case of a single oscillator, we find somewhat counterintuitively that the main source of decoherence is mechanical damping, despite the fact that the mechanical decay rates are three orders of magnitude slower than the cavity decay rate $\kappa$ in that example.  Again, we attribute this result to the coherent driving of the optical field that reduces its effective decoherence rate. We also find that the localized wave functions oscillate around the local minima of the effective potential with a frequency
\begin{equation}
\Omega_m \approx \omega_m -\frac{\kappa^2\omega_m^2}{8|g_0||\eta|^2}.
\end{equation}
The origin of these oscillations is the cavity fluctuations that act as a source of multiplicative noise on the mechanic, as can be seen from Eq. (\ref{qforce}). The cavity noise terms are multiplied with the position of the oscillator, and the oscillator thus favors to have small displacement. 

\section{Conclusion}

In summary, we have investigated a number of aspects of the quantum dynamics of multiple mechanical modes coupled to a single quantized cavity field mode via linear or quadratic optomechanical interactions in the single-photon strong coupling regime, where the single-photon optomechanical coupling coefficient is comparable to the mechanical frequency and the cavity decay rate $\kappa$ is large enough that the optical field can be adiabatically eliminated. At the mean-field level the coherent part of the cavity field provides an effective potential for the mechanics, revealing mechanical bistability for the case of a single mode and linear interactions, and a sub-critical bifurcation and associated double-well potential for the case of quadratic interactions. In addition to sideband cooling effects that are not addressed in this paper, a number of aspects of the resulting nonlinear dynamics have been investigated in the past in the classical regime, for instance the radiation pressure induced bistability of the mechanics that was already demonstrated nearly 30 years ago~\cite{classical_bistability1}.  A key finding here is that in the single-photon strong coupling regime of optomechanics, quantum noise changes the mean-field picture significantly, particularly since the noise component associated with the optical field is multiplicative.  Specifically, for a single mode and linear interactions we found the disappearance of mechanical bistability, reminiscent of the familiar cavity QED case, whereas for quadratic interactions we elucidated a quantum interference phenomenon based on bifurcation-induced wave-packet splitting.  For the case of two mechanical modes and linear interactions we explored how an initial entangled state, taken as the quantum mechanical ground state of the effective potential, evolves in the presence of optical and mechanical decay, finding that the mutual quantum information can persist for long times due to the fact that the mechanical modes interact with a common cavity mode.  Similar persistence of quantum effects for times orders of magnitude longer than the short cavity decay time $\kappa^{-1}$ were also found for the case of quadratic interactions, which is a plus for the possibility of observing these effects.  Moreover, these results should serve as a cautionary tale and a warning against making superficial order of magnitude arguments to ignore the effects of mechanical damping rate $\gamma$ for $\gamma << \kappa$.  Coherent driving of one or the other system can change things dramatically, and particular care must be taken under such conditions.

Clearly, this paper has only scratched the surface of the wealth of dynamical effects that can take place in multimode optomechanical systems, both in the classical and the quantum regimes. For example, nonlinearities need not be associated with the interaction of mechanical modes with a common optical field, but can also result from a number of other coupling mechanics, such as perhaps interacting with the phonon bath of a common substrate, providing e.g. interesting options for band structure engineering. Larger multimode systems also open interesting venues for the study of lattice systems, and multimode systems are also at the core of propagation studies, with interesting potential for quantum acoustics for example. This is an extremely rich area of investigation that we will continue to explore in forthcoming research.

\section{Acknowledgement}
This work is supported in part by the US National Science Foundation, The USA Army Research Office, and  the DARPA ORCHID and QuASAR programs through grants from AFOSR and ARO.


\begin{thebibliography}{10}
\bibitem{footnote0} The optical bistability considered in these experiments resulted from the dipole coupling of the light field to the internal  degrees of freedom of the atoms, and it was sufficient to treat their center-of-mass motion classically.
\bibitem{footnote}That same trend is seen for the larger mechanical decay rate but the asymptotic decay rate is bigger by an additive factor of $10^{-3}\kappa$.
\end{thebibliography}
\end{document}